\documentclass[12pt]{article}
\pdfoutput=1
\usepackage{graphicx}
\usepackage{epsfig}
\usepackage{amsmath,amssymb,hyperref,enumerate}
\usepackage{pstricks}
\usepackage{slashed}
\usepackage{amsfonts}
\usepackage{url}
\usepackage{color}
\usepackage{appendix}
\usepackage{bm}
\usepackage{subfig}
\usepackage{cite}
\hypersetup{colorlinks,citecolor= black,linkcolor= black}
\definecolor{nicered}{rgb}{0.7,0.1,0.1}
\definecolor{nicegreen}{rgb}{0.1,0.5,0.1}
\usepackage{setspace}

\allowdisplaybreaks

\def\({\left(}
\def\){\right)}
\def\[{\left[}
\def\]{\right]}

\usepackage[margin=1.in]{geometry}

\begin{document}

%\setlength{\baselineskip}{18pt}

%\singlespacing
\onehalfspacing
%\doublespacing

\begin{flushright}
OSU-HEP-14-04\\
UMD-PP-014-008
\end{flushright}

\renewcommand{\thefootnote}{\fnsymbol{footnote}}

%\vspace*{-0.5cm}
\begin{center}
\LARGE {\bf \Large 7 keV Scalar Dark Matter and the \\[-0.1in]
Anomalous Galactic X-ray Spectrum}
\end{center}

\vspace{0.5cm}

\begin{center}
{\large  K. S. Babu$^a$\footnote{Email:
babu@okstate.edu} and Rabindra N. Mohapatra$^b$\footnote{Email: rmohapat@physics.umd.edu}}

\end{center}

\begin{center}
\it $^a$Department of Physics, Oklahoma State University,
Stillwater, Oklahoma 74078, USA

\bigskip
\it $^b$Maryland Center for Fundamental Physics and Department of Physics\\
 University of Maryland, College Park, MD 20742, USA
\end{center}

\renewcommand{\thefootnote}{\arabic{footnote}}
\setcounter{footnote}{0}

\bigskip

\begin{abstract}

We present a simple model for a 7 keV scalar dark matter particle which also explains the recently
reported anomalous peak in the galactic X-ray spectrum at 3.55 keV in terms of its two photon decay.
The model is arguably the simplest extension of the Standard Model, with the addition of a real
scalar gauge singlet field subject to a reflection symmetry.  This symmetry breaks spontaneously at a
temperature of order few GeV which triggers the decay of the dark matter particle into two photons.  In this framework,
the Higgs boson of the Standard Model is also the source of dark matter in the Universe.  The model
fits the  relic dark matter abundance and the partial lifetime for two photon decay, while being consistent
with constraints from domain wall formation and dark matter self-interactions.  We show that all these
features of the model are preserved in its natural embedding into a simple dark $U(1)$ gauge
theory with a Higgs mechanism. The properties of the dark photon get determined in such a scenario. High precision
cosmological measurements can potentially test these models, as there are residual effects from domain wall
formation and non-negligible self-interactions of dark matter.

\end{abstract}

\newpage

\section{Introduction}

Recently two independent groups have reported the observation of a peak in the galactic X-ray spectrum at 3.55 keV~\cite{Bulbul:2014sua,Boyarsky:2014jta}, which cannot be
explained in terms of known physics and astrophysics.  If confirmed, this discovery will have a major impact in astrophysics, cosmology and
particle physics.  A natural interpretation of this result is in terms of the decay of a new particle with mass of 7.1 keV
into photons.  This new particle should be around today for its decay to be observed, which naturally
suggests that the decaying particle is also the dark matter of the Universe.  Such a warm dark matter particle
in this mass range~\cite{scott} has been known to fit other cosmological observations quite well, in some instances even better than
the canonical cold dark matter scenario.

While a sterile neutrino of mass 7.1 keV is an attractive explanation~\cite{Abazajian:2014gza} of this result in view of the measured small neutrino masses,
any model of this type
should also explain the observed partial lifetime for its decay into photons.
The axion which solves the strong CP problem should have mass well below an eV to be consistent with other astrophysical
observations, and thus will not fit the anomalous galactic X-ray spectrum,
although more general axion-like particles may do the required job~\cite{axion}.
It will be desirable to construct a variety of self-consistent models~\cite{others} that explain the cosmological data as well as
the galactic X-ray spectrum and confront them with experiments.

In this paper we propose a very simple model for warm dark that explains the 3.55 keV peak in the galactic X-ray spectrum.
Unlike the sterile neutrino, which is a fermion, and axion-like particles or the Majoron \cite{seckel},
which are pseudoscalars, our dark matter particle
is a real scalar.  The model is perhaps the simplest imaginable extension of the Standard Model, as it has the fewest number
of degrees of freedom added, namely one, corresponding to a real scalar field $\Phi$.  The Lagrangian of this scalar field
obeys a reflection symmetry $R$ under which  $\Phi \rightarrow -\Phi$,
which guarantees its longevity at cosmological time scales.  Such an extension of the Standard Model has been discussed
in the literature~\cite{real}, but our model differs in two crucial respects from previous discussions~\cite{real,real1}: (i) The mass of the dark matter
particle is in the keV range in our case, and (ii) The symmetry $R$ is broken spontaneously at an energy scale
of few GeV.  The breaking of $R$ allows for very small mixing between $\Phi$ and the Standard Model Higgs field $H$.  It is through
this mixing that $\Phi \rightarrow \gamma \gamma$ arises, which will turn out to be the leading decay mode of $\Phi$.  It is believed
that the SM Higgs boson $H$ is responsible for generation of masses of all particles; in our scenario $H$ is also
responsible for the origin of dark matter in the Universe, since $H \rightarrow \Phi \Phi$ produces the dark matter
particle in the early Universe at temperatures comparable to $M_H \simeq 126$ GeV.  When produced, the decay products never
thermalize, owing to the smallness of the mixing term.  The  momentum of $\Phi$ redshifts as the Universe cools, and
as we show here, leads to the correct abundance of relic dark matter for a reasonable choice of the few parameters of the model.
High precision cosmological and astrophysical observations can test this model, through imprints of domain wall formation
which occurs when the discrete symmetry $R$ breaks at a few GeV, and through non-negligible self-interaction of dark matter.\footnote{Although
the discrete symmetry $R$ is broken spontaneously at temperatures below a few GeV, it will turn out that the vacuum expectation
value of the field $\Phi$ is of order a few MeV.}

There is a natural embedding of the model presented here into a hidden $U(1)$ gauge theory where a
Higgs mechanism breaks the gauge symmetry.  The field $\Phi$ is identified in this case as the left-over Higgs scalar.  The resulting model is
perhaps the simplest gauge extension of the Standard Model.  In this case the properties of the dark photon (also referred to
as the hidden photon) get determined
from a fit to the  relic abundance of scalar dark matter and the observed X-ray anomaly.  We find that the dark photon must
have a mass in the range of 10 keV to a few MeV, the hidden $U(1)$ gauge coupling should naturally be of order 0.1, and
the kinetic mixing  of the dark photon with the photon and the $Z$ boson should be extremely small.

In Section 2 we describe our model with a real scalar field $\Phi$.
Section 3 discusses the cosmological implications of the model, including the anomalous
peak in the galactic X-ray spectrum, stability of the dark matter, and its relic abundance. Restoration of the
discrete reflection symmetry $R$ at temperature $T \sim$ few GeV and constraints from domain wall
formation are also discussed here, along with restrictions from dark matter self-interactions.  We
also discuss the consequences of the model if the discrete symmetry is relaxed.  In Section 4
we present an embedding of the model into a dark $U(1)$ gauge theory.  There we present constraint on the properties
of the dark photon.  Section 5 has our conclusions.

\section{Model}

The model is a very simple extension of the Standard Model (SM), with the addition of a real scalar gauge singlet field $\Phi$.
This field is subject a reflection symmetry $R$, under which $\Phi \rightarrow -\Phi$, with all other fields being invariant.
Note that this is the most economical extension of the SM, as it adds only a single new degree of freedom.

The Higgs potential of the model is given by
\begin{equation}
V = \frac{\lambda_H}{4} \left(H^\dagger H-v^2\right)^2 + \frac{\lambda_\Phi}{4} \left(\Phi^2-u^2\right)^2 + \frac{\lambda_{H\Phi}}{2}
\left(H^\dagger H-v^2\right)\left(\Phi^2-u^2\right)~.
\label{pot}
\end{equation}
Here $H$ is the SM Higgs doublet field, and the reflection symmetry $R$ has been applied.  Boundedness of the potential
requires the following conditions on the quartic couplings:
\begin{equation}
\lambda_H > 0,~~~\lambda_\Phi > 0,~~~ \lambda_{H\Phi} > -\sqrt{ 2 \lambda_H \lambda_\Phi}~.
\end{equation}
We shall see that the sign of the coupling $\lambda_{H\Phi}$ can be either positive or negative.  For positive $\lambda_{H\Phi}$ the
discrete symmetry $R$ will be shown to be restored above $T \sim 15$ GeV, while for negative $\lambda_{H\Phi}$ the symmetry $R$ will
never be restored.  As the Universe cools down, a phase transition occurs at $T \sim 15$ GeV in the case of positive $\lambda_{H\Phi}$,
which would lead to domain wall formation, which is studied and shown to be consistent in Sec. \ref{domain}.

The vacuum expectation values of
the neutral component of $H$ is denoted as $v$ and that of $\Phi$ is denoted $u$.  With this choice $\left\langle V \right\rangle =0$, which
is desirable from cosmology.\footnote{A constant has been added to the potential so that $\left\langle V \right\rangle =0$ is realized.
Since $v\simeq 174$ GeV and $u\sim$ a few MeV, the associated vacuum energy will be too large for a consistent evolution of the Universe, unless
it is cancelled by the addition of such a constant.}
Making an expansion $H = h/\sqrt{2}+v$ and $\Phi=\phi+u$, with $v \simeq 174$ GeV, we arrive at the $2 \times 2$ mass matrix for the
$h-\phi$ system:
\begin{eqnarray}
{\cal M}^2 = \left(\begin{matrix}\lambda_H v^2 & \sqrt{2}\, \lambda_{H\Phi}\, u v \\
\sqrt{2}\, \lambda_{H\Phi}\, u v & 2 \lambda_{\Phi} u^2 \end{matrix}\right)~.
\end{eqnarray}
We shall be interested in the limit where $\lambda_{H\Phi} \ll 1$, which is needed for the stability of the dark matter.
Furthermore, in order to explain the galactic X-ray anomaly, the mass of the dark matter field should be 7.1 keV, which
is much lighter than $M_h \simeq 126$ GeV.  While the coupling $\lambda_{\Phi}$ will turn out to be small, it will obey
$\lambda_{\Phi} \gg \lambda_{H\Phi}^2/\lambda_H$.  Thus we obtain the mass eigenvalues to an excellent approximation as
\begin{eqnarray}
M_h^2 = \lambda_H v^2, \nonumber \\
M_\phi^2 = 2 \lambda_\Phi u^2,
\end{eqnarray}
along with the $h-\phi$ mixing angle
\begin{equation}
\tan 2\theta_{H\Phi} =\frac {2 \sqrt{2} \lambda_{H\Phi}\, u v}{\lambda_H v^2-2 \lambda_\Phi u^2} \simeq \frac{2 \sqrt{2} \lambda_{H\Phi}\, uv}
{M_h^2}~.
\label{mixing}
\end{equation}
We identify $M_\phi \simeq 7.1$ keV and $M_h \simeq 126$ GeV, corresponding to the masses of the dark matter particle and the SM Higgs boson
respectively.  Note that in the
limit $\lambda_{H\Phi} \rightarrow 0$, the $\phi$ field would decouple from all of the SM particles.  For nonzero values of
$\lambda_{H\Phi}$, $\phi$ will interact with other SM fields via its mixing with $h$.  In addition, the $\lambda_{H\Phi}$ coupling
will lead to the decay $h \rightarrow \phi \phi$ as well as scattering processes such as $hh \rightarrow \phi \phi$,
$W^+ W^- \rightarrow \phi \phi$, $ZZ \rightarrow \phi \phi$ and $t \overline{t} \rightarrow \phi \phi$, which
will be the source of dark matter production in the early Universe.  We shall see that the decay $h \rightarrow \phi \phi$ is the
dominant source of dark matter.
We shall insist that  $h \rightarrow \phi \phi$ decay be out of thermal equilibrium at all epochs, which will set an
upper limit on the coupling $|\lambda_{H\Phi}| < 7 \times 10^{-8}$.  By demanding that the abundance of $\phi$ today fits the
relic abundance of dark matter, we determine the value of $|\lambda_{H\Phi}| \simeq 4.7 \times 10^{-9}$, which is somewhat
smaller than the upper limit obtained by demanding that $\phi$ never thermalizes.

The new parameters of the model will be determined to be $|\lambda_{H\Phi}| \sim 4.7 \times 10^{-9}$, $\lambda_\Phi \sim 2 \times 10^{-7}$
and $|u| \sim (3-8)$ MeV.  The value of $\lambda_{\Phi}$ is in the interesting range for a 7 keV scalar dark matter to have
self-interactions that may be observable in astrophysical settings.
It should be noted that such small values of the quartic scalar couplings are technically natural, as the other
larger couplings of the model do not induce these quartic couplings by themselves.

We reemphasize that the main new ingredient in our model compared to the existing ones~\cite{real,real1} with a real singlet field added to SM is that our scalar field acquires a non-zero vacuum expectation value which makes it relevant for the 3.5 keV X-ray lines. This leads to a completely different phenomenology, developed in the next section, compared to those of Ref. \cite{real,real1}.

\section{Cosmological and astrophysical implications of the model}

In this section we shall discuss the various implications of the model for cosmology and astrophysics.
The anomalous X-ray peak in the galactic spectrum
and the relic abundance of dark matter particle $\phi$ will be our primary focus.
We shall also establish that the dark matter is adequately stable.
Improved cosmological observations will be shown to  test  the model, in the residual effects
on Cosmic Microwave Background (CMB) anisotropy arising from low temperature domain walls that are formed in the model,
and in observable effects of the self-interactions of the dark matter particle.

\subsection{Galactic X-ray spectrum and \boldmath{$\phi \rightarrow \gamma \gamma$} decay}

As noted earlier, the dark matter particle $\phi$ has a small admixture of the SM Higgs boson $h$. We denote the mass eigenstates
as simply $h$ and $\phi$. This mixing will cause the decay
$\phi \rightarrow \gamma \gamma$.  All other decays are kinematically forbidden, except for the decay $\phi \rightarrow \nu \nu$, which
we shall discuss in the next subsection. The decay rate for $\phi \rightarrow \gamma \gamma$ is given by
\begin{equation}
\Gamma(\phi \rightarrow \gamma \gamma) = \left(\frac{\alpha}{4 \pi}\right)^2 |F|^2 \,\sin^2\theta_{H\Phi}\,\frac{ G_F m_\phi^3}{8 \sqrt{2} \pi}
\end{equation}
where
\begin{equation}
F = F_W(\beta_W) + \sum_fN_c Q_f^2 F_f(\beta_f)
\end{equation}
with
\begin{eqnarray}
\beta_W &=& \frac{4 M_W^2}{M_\phi^2},~~\beta_f = \frac{4 M_f^2}{M_\phi^2},\nonumber \\
F_W(\beta) &=& 2+3\beta+3 \beta(2-\beta) f(\beta),\nonumber\\
F_f(\beta) &=& -2\beta[1+(1-\beta)f(\beta)],\nonumber\\
f(\beta) &=& \arcsin^2[\beta^{-1/2}]~.
\end{eqnarray}
Here $Q_f$ is the electric charge of the fermion and $N_c$ is the color degrees of freedom.
Since $\beta_W \gg 1$ and $\beta_f \gg 1$ for all charged fermions in the SM, $F$ can be
approximated as $F = 7 - (4/3)(8) = -11/3$, where $F_W = 7$ and $F_f = -4/3$ and
the factor 8 arises from the summation of $N_c Q_f^2$ over all charged fermions
of the SM.

For $M_\phi = 7.1$ keV we find
\begin{equation}
\Gamma(\phi \rightarrow \gamma \gamma) = 6.2 \times 10^{-33} \sin^2\theta_{H\Phi} ~{\rm GeV}.
\end{equation}
The galactic X-ray excess corresponds to a inverse rate of \cite{Bulbul:2014sua,Boyarsky:2014jta}
\begin{equation}
\Gamma^{-1}(\phi \rightarrow \gamma \gamma) = (4 \times 10^{27}\, {\rm s} - 4 \times 10^{28}\, {\rm s}),
\end{equation}
which leads to a determination of the range of mixing angle
\begin{equation}
|\sin\theta_{H\Phi}| = (5.6 \times 10^{-13} - 1.8 \times 10^{-13})~.
\label{mix1}
\end{equation}
Using Eq. (\ref{mixing}), we can arrive at the range for $|\lambda_{H\Phi}\, u|$ to be
\begin{equation}
|\lambda_{H\Phi}\, u| = (3.6 \times 10^{-11} - 1.2 \times 10^{-11})~{\rm GeV}~.
\label{range}
\end{equation}

\subsection{Other decays of \boldmath{$\phi$}}

Neutrinos and photons are the only particles of the SM that are lighter than $\phi$.  We should check if
the decay $\phi \rightarrow \nu \nu$ is sufficiently long-lived for $\phi$ to constitute the dark matter.
For this purpose we assume that the neutrino masses are of Majorana type and the effective Lagrangian for
neutrino masses is given by
\begin{equation}
{\cal L}_\nu^{\rm eff} = \frac{\kappa}{2} (LL)(HH)
\end{equation}
with $\kappa$ having inverse dimensions of mass and $L$ being the SM lepton doublet. The neutrino mass matrix is then $M_\nu = \kappa v^2$.
Noting that the $H$ field has a small admixture of $\phi$ as given in Eq. (\ref{mixing}), we obtain the
decay rate for $\phi \rightarrow \nu \nu$ to be
\begin{equation}
\sum_\nu \{\Gamma(\phi \rightarrow \nu\nu) + \Gamma(\phi \rightarrow \overline{\nu}\, \overline{\nu}\} =
\frac{{\rm Tr}(M_\nu^\dagger M_\nu)}{8 \pi v^2}\sin^2\theta_{H\Phi} M_\phi~.
\end{equation}
If we use a hierarchical neutrino mass spectrum with the largest mass being $0.049$ eV, we obtain the
lifetime to be
\begin{equation}
\left(\sum_\nu \{\Gamma(\phi \rightarrow \nu\nu) + \Gamma(\phi \rightarrow \overline{\nu} \,\overline{\nu}\}\right)^{-1} =
(9 \times 10^{31} - 8.8 \times 10^{32})~ {\rm s}
\label{lifetime}
\end{equation}
corresponding to $\sin\theta_{H\Phi}$ given in Eq. (\ref{mix1}).  Even if the neutrino masses are
quasi-degenerate with the mass of each species being 0.5 eV, this lifetime will be shortened only by a factor
of 300 relative to Eq. (\ref{lifetime}).  Thus $\phi \rightarrow \gamma \gamma$ is the more dominant
decay, and $\phi$ is rather stable at cosmological timescales.

\subsection{Relic abundance of dark matter}
\label{relic}

In addressing the relic abundance of dark matter $\phi$, we consider the various sources of production of dark matter $\phi$
in the early universe. We assume that at high temperatures there is no source of $\phi$ production (such as from inflaton decay).
The Standard Model Higgs boson is the primary source of $\phi$ production in our scenario.  This can occur in two different ways.
(i) via Higgs boson decay $h \rightarrow \phi \phi$, and (ii) via scattering of two SM particles by exchanging $h$ to produce
a pair of $\phi$'s.
\begin{figure}[h!]
\centering
\includegraphics[scale=0.5]{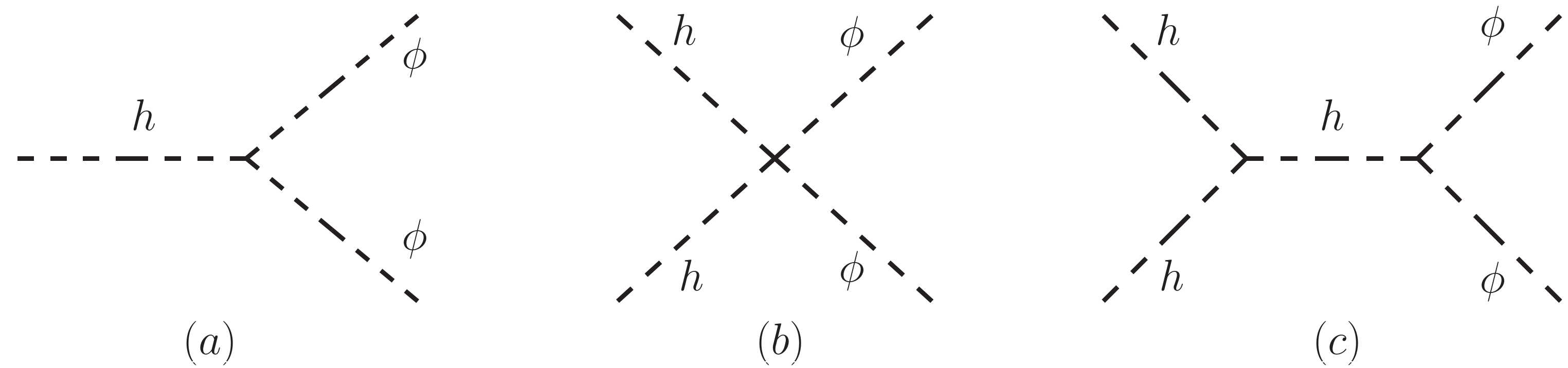}\\[0.1in]
\includegraphics[scale=0.5]{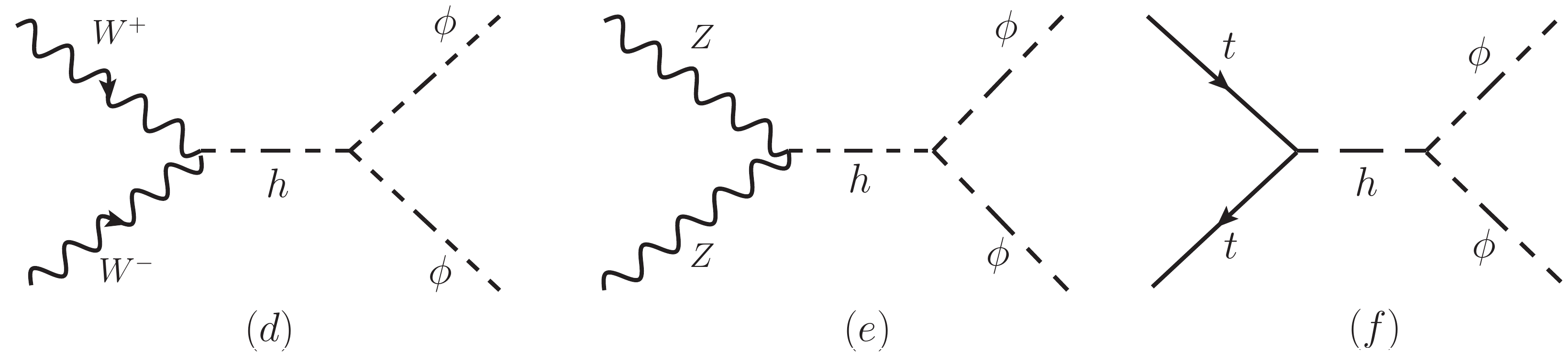}
\caption{Diagrams for $\phi$ production.  (a): $h \rightarrow \phi \phi$, (b),(c): $hh \rightarrow \phi \phi$, (d): $W^+ W^- \rightarrow
\phi \phi$, (e): $ZZ \rightarrow \phi \phi$, and (f): $t \overline{t} \rightarrow \phi \phi$.}
\label{feyn}
\end{figure}
The relevant scattering processes are  $h h \rightarrow \phi \phi$, $W^+ W^- \rightarrow \phi \phi$, $ZZ \rightarrow \phi \phi$ and
$t\overline{t} \rightarrow \phi \phi$.
Both the decay and the scattering processes
occur through the mixed $H-\Phi$ quartic coupling $\lambda_{H\Phi}$.  Once the $\phi$ particles are produced,
we require that they do not thermalize with the rest of the plasma.  This restriction
can be inferred from the decay rate and the annihilation cross sections for $\phi$ production shown in Fig. \ref{feyn}.
The decay rate is given by
\begin{equation}
\Gamma(h \rightarrow \phi \phi) = \frac{|\lambda_{H\Phi}|^2}{16 \pi}\frac{v^2}{M_h}~,
\end{equation}
with $v\simeq 174$ GeV.
The (unpolarized and spin-averaged) annihilation cross sections are given by
\begin{eqnarray}
\sigma(h h \rightarrow \phi \phi) &=&  \frac{|\lambda_{H\Phi}|^2}{32 \pi s}
\frac{\left(s+2M_h^2\right)^2} {\sqrt{1-4M_h^2/s}~(s-M_h^2)^2}~,\nonumber \\
\sigma(VV \rightarrow \phi \phi) &=& \frac{|\lambda_{H\Phi}|^2 }{72 \pi s}
\frac{(3M_V^4 - s M_V^2 + s^2/4)} {\sqrt{1-4M_V^2/s}~(s-M_h^2)^2}~, \nonumber \\
\sigma(t\overline{t} \rightarrow \phi \phi) &=& \frac{3 |\lambda_{H\Phi}|^2}{64 \pi} \frac{M_t^2}{\sqrt{1-4M_t^2/s}~(s-M_h^2)^2}~.
\end{eqnarray}
Here $VV$ stands for either $W^+W^-$ or $ZZ$, the factor 3 in $\sigma(t \overline{t} \rightarrow \phi \phi$) arises from color
summation, and $M_h \simeq 126$ GeV is the mass of the SM Higgs boson exchanged in these processes (see Fig. \ref{feyn}).

If the decay rate and the annihilation rate $\left\langle\sigma n |v|\right\rangle$ are
much smaller than the Hubble expansion rate at temperature of order $M_h$, the dark matter particle $\phi$ will never be in thermal
equilibrium.  The Hubble expansion rate at temperature $T$ is given as
\begin{equation}
H(T) = 1.66 {g^*}^{1/2} \frac{T^2}{M_{\rm Pl}}~.
\end{equation}
Here $g^*$  is the effective number of relativistic degrees of freedom at $T$.
The value of $g^*$ at $T \sim M_h$ is 389/4, which accounts for all the SM particles minus the top quark, and adds one degree for $\phi$.
Demanding that $\Gamma(h \rightarrow \phi\phi) < H(T \sim M_h)$ gives the constraint $|\lambda_{H\Phi}| < 6.7 \times 10^{-8}$.
The constraint from the annihilation rates are much weaker.  For example,
the condition $\left\langle\sigma(hh\rightarrow \phi \phi) \,n\, |v|\right\rangle < H(T \sim M_h)$ leads to
$|\lambda_{H\Phi}| < 2.4 \times 10^{-6}$, where $n \simeq (0.1) \,T^3$ is used for the equilibrium number density of the Higgs boson $h$,
along with  the relations $s=4E^2$ and $E = 3.15\, T$. We shall require the stronger of the constraints.

It should be noted
that the decay $h \rightarrow \phi \phi$ as well as the scattering processes shown in Fig. \ref{feyn} (except for  \ref{feyn} (b))
occur only below the electroweak phase transition temperature $T_{\rm EW} \simeq 153$ GeV,
as the scalar interaction vertex, $h \rightarrow \phi \phi$  develops only in this regime. Nevertheless, these processes occurring in the temperature range below 153 GeV and above 5 GeV (below 5 GeV the $h$, $W^\pm$ and $Z$ bosons go out of thermal equilibrium) will play a crucial role in the creation of dark matter.

To evaluate the relic density of dark matter $\phi$, we solve the Boltzmann equation for the evolution of the number density $n_\phi$ of $\phi$.
It is given as
\begin{equation}
 \dot{n}_\phi + 3 H n_\phi =  -\left\langle \Gamma(h \rightarrow \phi \phi \right\rangle (n_\phi-n_0^h)
 -\sum_{i=h,W,Z,t} \left\langle \sigma_i(i \bar{i} \rightarrow \phi \phi) |v| \right\rangle (n_\phi^2-(n^i_0)^2)~.
\label{bolt}
\end{equation}
Here $n_0^i$ is the equilibrium number density of species $i$.  We shall use Boltzmann distribution functions for both fermions
and bosons, which is a good approximation.  For the range of $|\lambda_{H\Phi}| < 6.7 \times 10^{-8}$, $\phi$ number density
will never be significant, and thus we can set $n_\phi = 0$ on the RHS of Eq. (\ref{bolt}).  That is, we can ignore back reactions of the
type $\phi \phi \rightarrow hh$.  The $\left\langle~ \right\rangle$ symbol stands for thermal averaging.  Changing variable from time $t$
to temperature $T$ (using conservation of entropy -- $RT$ = constant), and defining
\begin{equation}
z = \frac{M_h}{T},~~~ ~ f_\phi = \frac{n_\phi}{T^3},~~~~ f_0^i = \frac{n_0^i}{T^3},~~~K = \frac{1.66\, (g^*)^{1/2} }{M_{\rm Pl}}~,
\end{equation}
Eq. (\ref{bolt}) can be recast (taking $n_\phi = 0$ on the RHS) as
\begin{equation}
\frac{df_\phi}{dz} = \frac{\left\langle \Gamma(h \rightarrow \phi \phi)\right\rangle}{K M_h^2} z f_0^h +
\sum_{i=h,W,Z,t}\frac{\left\langle \sigma(i \bar{i} \rightarrow \phi \phi)|v|\right\rangle M_h}{K z^2} \left(f_0^i\right)^2
\label{bolt2}
\end{equation}
The region of validity of this equation is when species $i=h,W,Z,t$ is in
equilibrium (which occurs down to temperature $T$ about 15-20 times below $M_i$)
and when $\phi$ is not in equilibrium.  Since both in decay and in scattering two $\phi$'s are produced, the net abundance of $\phi$ should
be twice that obtained by solving this equation.  The thermal averages can be expressed as follows \cite{gondolo}:
\begin{eqnarray}
\left\langle \Gamma(h \rightarrow \phi \phi)\right \rangle &=& \Gamma(h \rightarrow \phi \phi) \frac{K_1(z)}{K_2(z)}, \\
\left\langle \sigma(hh\rightarrow \phi \phi)|v|\right\rangle &=& \frac{1}{8 M_h^4 T K_2^2\left(\frac{M_h}{T}\right)}\int_{4M_h^2}^\infty \,\sigma\,
(s-4M_h^2)\,\sqrt{s}\,K_1\left(\frac{\sqrt{s}}{T}\right)\,ds
\end{eqnarray}
and similarly for other scattering processes.  Here $K_{1,2}$ are modified Bessel functions.

Defining
\begin{equation}
f_\phi = \frac{|\lambda_{H\Phi}|^2 }{16 \pi}\, \frac{v^2}{K M_h^3}\, \kappa_\phi,
\end{equation}
the Boltzmann equation for $\kappa_\phi$ can be written down as
\begin{equation}
\frac{d\kappa_\phi}{dz} = \frac{K_1(z)}{K_2(z)} z f_0^h(z) + \sum_{i=h,V,t}c_i\left\{f_0^i(x_iz)\right\}^2 \frac{1}{K_2^2(x_iz)}
\int_{4z^2}^{\infty} dy \,K_1(x_i\sqrt{y})\,\frac{\sqrt{y-4 z^2}}{(y-z^2)^2}\,h_i(y)~.
\label{boltz2}
\end{equation}
Here we have defined
\begin{eqnarray}
x_i &=& \frac{M_i}{M_h},~~c_h = \frac{M_h^2}{16 v^2},~~c_V = \frac{M_V^3}{36 M_h v^2},~~c_t = \frac{3}{32}\frac{M_t^7}{M_h^5 v^2}, \nonumber \\
 f_0^i(x_iz) &=& \frac{g_i}{2\pi^2}\int_0^\infty y^2 e^{-\sqrt{y^2 + x_i^2 z^2}} dy,\nonumber \\
 h_h &=& (2+y)^2,~~h_V= 3-yz^2 x_V^4+\frac{y^2z^4 x_V^8}{4},~~h_t = y z^2,
\end{eqnarray}
where the summation over $V$ includes $V=W,Z$, $x_i = 1$ for $i=h$, and $g_i$ account for the spin degrees with
$g_h = 1,\,g_V=3$ and $g_t=2$.

\begin{figure}[h!]
\centering
\includegraphics[scale=0.5]{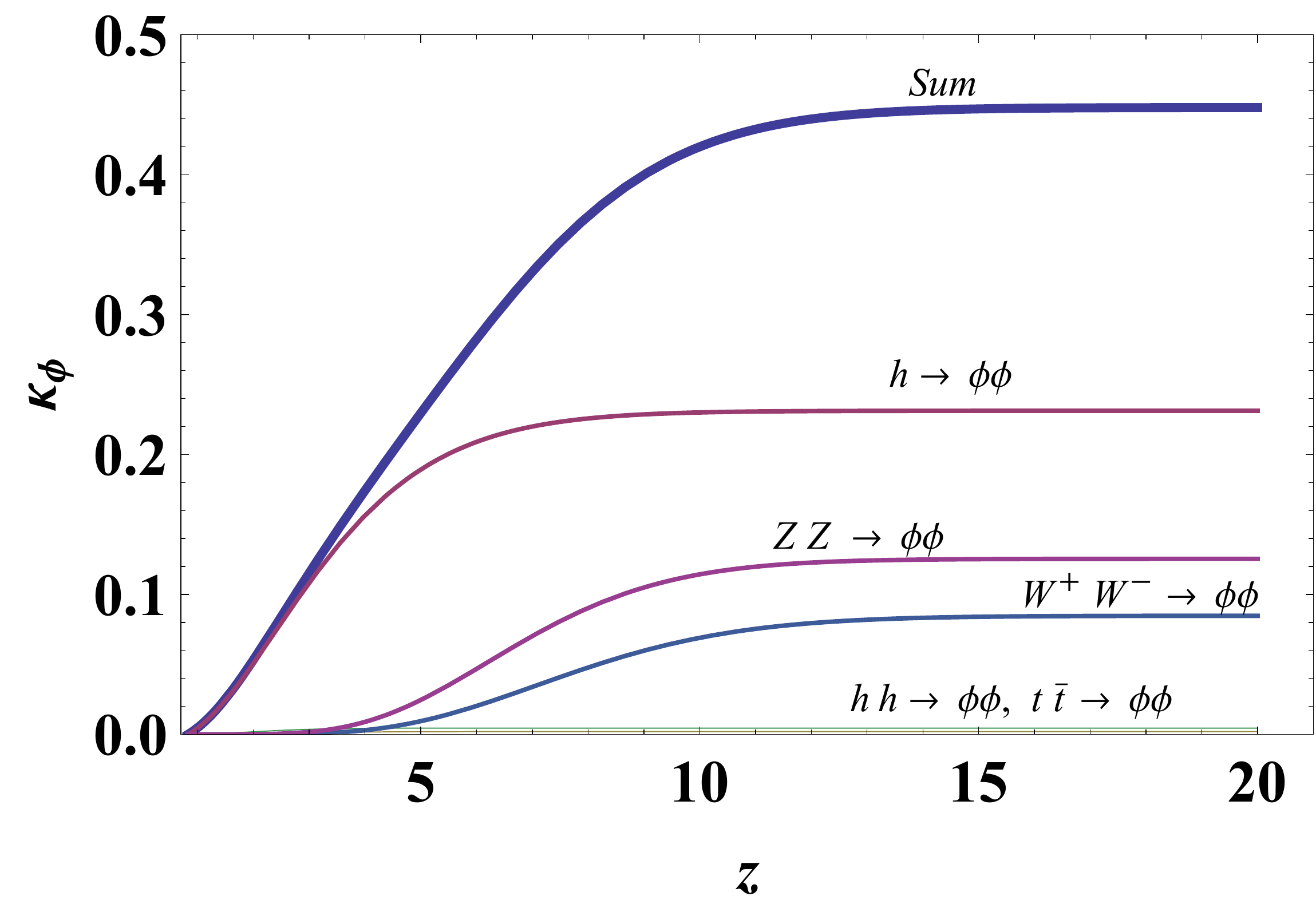}
\caption{Results for $\kappa_\phi$ obtained by numerical integration of Eq. (\ref{boltz2}). Contributions from various subprocesses as well as
the sum are shown.}
\label{kappa}
\end{figure}

We have integrated Eq. (\ref{boltz2}) numerically from $z=0.83$ to $z=20$ (corresponding to the electroweak phase transition temperature
and the freeze-out temperature of the fields $h, W,Z$ and $t$) with the boundary condition $\kappa_\phi =0$ at $z=0.83$.  The results are
shown in Fig. \ref{kappa}.  The leading source of $\phi$ generation is $h \rightarrow \phi \phi$ decay, contributing to $\kappa_\phi = 0.231$
at $z=20$.  The contributions from the scattering processes to $\kappa_\phi$ are $0.125$ from $ZZ \rightarrow \phi \phi$,
$0.085$ from $W^+ W^- \rightarrow \phi \phi$, $0.005$ from $t\bar{t} \rightarrow \phi\phi$ and $0.002$ from $hh \rightarrow \phi \phi$.
The net asymptotic value of $\kappa_\phi$ is 0.445, which we shall use in our estimate of relic abundance of $\phi$.

Using the expression for the decay width for
$h \rightarrow \phi \phi$
we find the number density of $\phi$ at $T \sim M_h$ to be
\begin{equation}
\left(\frac{n_\phi}{T^3}\right)_{T\sim M_h} \approx \frac{0.89 \,|\lambda_{H\Phi}|^2}{16 \pi}\frac{v^2}{M^3_HK}~.
%+\frac{1.2\lambda^2_{H\phi}}{256\pi K m_h}f^2_0|_{T\sim M_h}
\end{equation}
A factor of 2 has been included to go from $\kappa_\phi =0.445$ to the value used here (0.89) owing to the
two $\phi$ fields produced in each decay.  We then find
\begin{equation}
\left(\frac{n_\phi}{T^3}\right)_{T\sim M_h} \approx 4.1 \times 10^{-3} \left(\frac{\lambda_{H\Phi}}{4.5 \times 10^{-9}}\right)^2~.
\end{equation}

The $\phi$ fields produced in the decay of $h$ never thermalize.  As the Universe cools, the momentum of $\phi$ redshifts.
The number density of $\phi$ today is then estimated to be
\begin{equation}
n_\phi^0 \approx 4.1 \times 10^{-3} \left(\frac{\lambda_{H\Phi}}{4.5 \times 10^{-9}}\right)^2 \frac{T_0^3}{\xi}
\label{density}
\end{equation}
where
\begin{equation}
\xi = \frac{g^*(T=M_h)}{g^*(T=0.1~{\rm MeV})} = \frac{389}{12} = 43.22~.
\end{equation}
The decoupling of various species heats the plasma, while the $\phi$ field is not affected.  The factor $\xi$ in Eq. (\ref{density})
accounts for the heating of the plasma. $T_0$ in Eq. (\ref{density}) if the present day temperature, $T_0 = 2.73^0$ K.  From Eq. (\ref{density})
with the use of $\rho_\phi = M_\phi n_\phi^0$ for the present density of $\phi$ and the definition $\Omega_\phi = \rho_\phi/\rho_{\rm crit}$
we obtain
\begin{equation}
\Omega_\phi h^2 \approx 0.11 \left(\frac{\lambda_{H\Phi}}{4.5 \times 10^{-9}}\right)^2 \left(\frac{43.22}{\xi}  \right)~.
\end{equation}
Thus we see that the desired value of $\Omega_\phi h^2 = 0.1198$ can be produced with $|\lambda_{H\Phi}| = 4.7 \times 10^{-9}$.
We can also determine the range of $u$ from Eq. (\ref{range}) to be
\begin{equation}
u = (7.7 - 2.6)~ {\rm MeV}~.
\label{uvalue}
\end{equation}

Since the $\phi$ particle is never in thermal equilibrium, it does not contribute significantly to big bang nucleosynthesis.
The model thus predicts $\Delta N_\nu = 0$ for BBN.

\subsection{Restoration of discrete symmetry and domain wall constraints}
\label{domain}

The reflection symmetry $R$ (under which $\Phi \rightarrow -\Phi$) is spontaneously broken in our model.  Could this symmetry be
restored at high temperature?  If so, domain walls would form as the Universe cools and undergoes a phase transition.  Here we
discuss this issue and show the consistency of the model.

\begin{figure}[h!]
\centering
\includegraphics[scale=0.5]{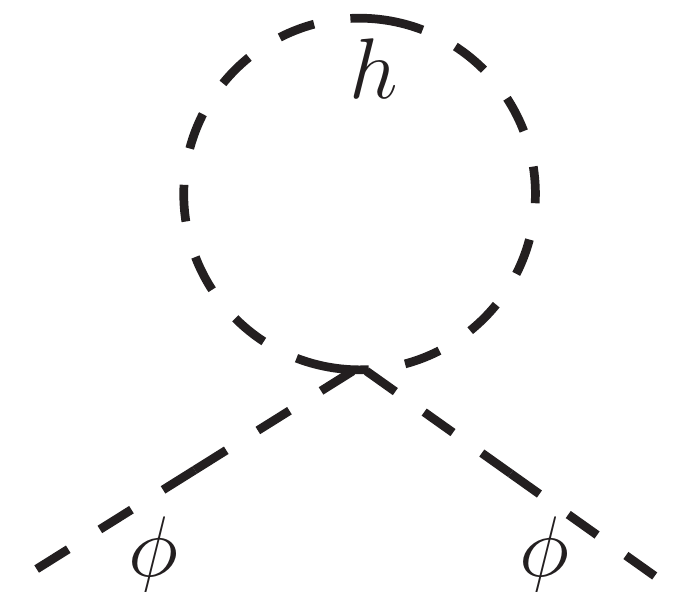}
\caption{Diagram generating finite temperature mass for $\phi$ }
\label{mass}
\end{figure}

Since the $\phi$ field is never in thermal equilibrium, $\phi \phi \rightarrow \phi \phi$ scattering process with a cross section proportional
to $|\lambda_\Phi|^2$ will not induce a temperature dependent mass for $\phi$.  However, the Higgs field $h$ has a thermal distribution,
and therefore the scattering of $\phi$ with the $h$ field will generate a temperature dependent mass for $\phi$.
The relevant diagram is shown in Fig. \ref{mass}.  The finite temperature mass of $\phi$ is found to be \cite{dolan}
\begin{equation}
M_\phi^2(T) = \frac{\lambda_{H\Phi}\,T^2}{4 \pi^2} \int_0^\infty dx \frac{x^2}{\sqrt{x^2+M_h^2/T^2}} \frac{1}{e^{\sqrt{x^2+M_h^2/T^2}}-1}~.
\label{temp}
\end{equation}
For $T \gg M_h$ and for $T \ll M_h$ this can be evaluated to be
\begin{eqnarray}
M^2_\phi(T) &\simeq& \left(\frac{\lambda_{H\Phi}}{24}\right) T^2,~~~T \gg M_h, \nonumber \\
M^2_\phi(T) & \simeq& \left(\frac{\lambda_{H\Phi}M_h T} {4 \pi^2}\right)\, K_1\left(\frac{M_h}{T}\right), ~~~T \ll M_h \nonumber \\
&\simeq& \frac{\lambda_{H\Phi}}{4 \sqrt{2} \pi^2} T^2
\sqrt{\frac{M_h}{T}\,}e^{-M_h/T} \, \left(1 + \frac{3}{8} \frac{T}{M_h} - \frac{15}{128} \frac{T^2}{M_h^2} + ...  \right)
\end{eqnarray}
For the parameters of our model, the approximation $T/M_h \ll 1$ is more relevant.  Note that from Eq. (\ref{pot}), there is a negative
mass term for $\phi$ field, given by $-\lambda_\Phi u^2$.  Combining this with the temperature dependent mass of Eq. (\ref{temp}), we find,
for positive $\lambda_{H\Phi}$, that  at $T = T_R$ a phase transition occurs which would restore the $R$ symmetry.  $T_R$ is obtained
by demanding that $M^2_\phi(T) - \lambda_\Phi u^2$ turns positive, which can happen only when $\lambda_{H\Phi}$ is positive.  In this case,
for $u = 7.7$ MeV, $\lambda_\Phi = 4.3 \times 10^{-7}$, and $\lambda_{H\Phi} = 4.7 \times 10^{-9}$, we find
\begin{equation}
T_R \simeq 15.2 ~{\rm GeV}~.
\end{equation}
For negative values of $\lambda_{H\Phi}$, which is allowed in the model, at high temperature there is no restoration of $R$ symmetry.

Since the reflection symmetry is restored at $T_R \simeq 15.2$ GeV for the case of positive $\lambda_{H\Phi}$,
domain wall formation will occur in the model in this case~\cite{okun}.
We proceed to show the consistency of the model with constraints from domain wall formation.
We note that there is no domain wall formation if $\lambda_{H\Phi}$ takes negative values.
It should be noted that the energy density in domain walls is controlled by the vacuum expectation value $u \sim$ few MeV, rather than
the phase transition temperature $T_R$.

At temperatures of order $T_R \sim$ a few GeV, the Higgs potential relevant is simply
\begin{equation}
V = \frac{\lambda_\Phi}{4} (\Phi^2-u^2)^2~.
\end{equation}
Classical field configurations with the boundary conditions $\Phi(\pm \infty) = \pm u$ then exist.  For propagation along the $z$ direction,
this solution obeys the following equations:
\begin{equation}
\frac{\partial \phi}{\partial t} = 0,~~~~\frac{\partial \phi}{\partial z} - \sqrt{2V(\phi)} = 0,
\end{equation}
\begin{equation}
E_{\rm min} = \int_{ \phi(-\infty)}^{\phi(\infty)} d \phi' \sqrt{2V(\phi')}~.
\end{equation}
The (kink) solution to this set of equations is
\begin{equation}
\phi = u \tanh\left(\sqrt\frac{\lambda_\Phi}{2}\, u z \right)
\end{equation}
with the energy per unit area $\sigma_{\rm kink}$ given by
\begin{equation}
\sigma_{\rm kink} = \frac{2 \sqrt{2}}{3} \sqrt{\lambda_\Phi}\,u^3~.
\end{equation}
The density in domain wall today is $\sigma_{\rm kink}/t_0$ with $t_0 \simeq 13.8 \times 10^9$ years being the present age
of the Universe.  If we demand $\Omega_{\rm domain~wall} = (\rho_{\rm domain ~ wall}/\rho_{\rm crit})$ to be less than one,
we obtain
\begin{equation}
\frac{2}{3} u^2 M_\phi \leq 5.4 \times 10^{-5} ~{\rm GeV}^3,
\end{equation}
or $|u| \leq 3.4$ GeV for $M_\phi = 7.1$ keV.

It is not enough, however, to simply demand that the domain wall does
not over-close the Universe.  The presence of domain wall can alter the CMB anisotropy significantly unless $\Omega_{\rm domain~wall}
\leq 10^{-5}$, in which case we obtain $|u| \leq 10.7$ MeV.
The model parameters allow for $|u| = (7.7-2.6)$ MeV, see Eq. (\ref{uvalue}).  Thus we see that the scenario presented is
consistent with domain wall constraints.  However, since the upper limit from CMB anisotropy is close to the prediction of
the model, precision CMB measurements can potentially unravel new contributions of this type. We expect contributions
to CMB anisotropy from domain walls at the level of $(0.52-0.06)\times 10^{-5}$. It would be desirable to
further study this question.

\subsection{Dark matter self-interaction constraints}

In our model, $\phi \phi \rightarrow \phi \phi$ interaction occurs through the quartic coupling $\lambda_\Phi$.  There are
rather severe constraints on self-interaction of dark matter from dense cores of galaxies and galaxy clusters where the
velocity distribution can be isotropized.  Constraints from such halo shapes, as well as from dynamics of bullet cluster
merger have been used to infer an upper limit on the dark matter self-interaction cross section.  A recent analysis \cite{self}
which includes the gravitational potential of both baryons and dark matter finds this cross section to be limited by
\begin{equation}
\frac{\sigma}{M_\phi} < 1\,{\rm barn}/{\rm GeV}~.
\end{equation}
For $M_\phi = 7.1$ keV, this translates into $\sigma < 7 \times 10^{-30}$ cm$^2$.  We take this to be a conservative upper
limit, although limits as strong as $\sigma/M_\phi < 0.1$ barn/GeV have been claimed.

 In our model the Lagrangian relevant for $\phi \phi \rightarrow \phi\phi$ scattering, which follows from Eq. (\ref{pot}) is
 \begin{equation}
 {\cal L}_{\phi \phi} = -\sqrt{\lambda_\Phi} \,M_\phi\, \phi^3 -\frac{\lambda_\Phi}{4} \, \phi^4~.
 \end{equation}
 The cross section for $\phi \phi \rightarrow \phi \phi$ scattering is found to be
\begin{eqnarray}
\sigma = \frac{9 \lambda_\Phi^2}{8 \pi s} \left[ 1 + \frac{9 M_\phi^4}{(M_\phi^2-s)^2} +
\frac{12 M_\phi^2 \,(2s - 3 M_\phi^2)}{(s-M_\phi^2)(s-3M_\phi^2)}
+
\frac{12 M_\phi^2\left(M_\phi^4 + 3 M_\phi^2 s - s^2\right)}{(s-M_\phi^2)(s-2M_\phi^2)(s-4M_\phi^2)}{\rm ln} \left(\frac{s-3M_\phi^2}{M_\phi^2}\right) \right].
\end{eqnarray}
In the nonrelativistic limit, $(s-4 M_\phi^2) \ll M_\phi^2$, this reduces to
\begin{equation}
\sigma \simeq \frac{9 \lambda_\Phi^2}{2 \pi M_\phi^2}\left[1 - \frac{19}{12} \left(\frac{s-4 M_\phi^2}{M_\phi^2}\right) + \frac{257}{144} \left(
\frac{s-4M_\phi^2}{M_\phi^2}\right)^2 + ..  \right]~.
\label{cross1}
\end{equation}
Demanding that $\sigma/M_\phi < 1$ barn/GeV, with $M_\phi = 7.1$ keV, we obtain from Eq. (\ref{cross1}) the limit
\begin{equation}
\lambda_{\Phi} < 8 \times 10^{-7}.
\end{equation}
Very interestingly, this is close to the limit derived from constraint on the VEV $u$ and the 7.1 keV mass of the dark
matter particle.  For $|u|=(7.7-2.6)$ MeV, we have $\lambda_\Phi = (4.2 \times 10^{-7}-3.7 \times 10^{-6})$.  The higher end
of $|u|$ is preferred here, which would make domain wall contributions to CMB in the more interesting range.  We see that
the model predicts dark matter self-interactions in the interesting range to play an important cosmological role.
In particular, the dark matter self interaction cross section is predicted to be $\sigma/M_\phi > 0.25$ barn in the model
with a spontaneously broken discrete symmetry.

\subsection{Supernova energy loss constraints}

The dark matter $\phi$ with a mass of 7.1 keV cannot be produced in stars such as the Sun (core temperature of about
2 keV) and in horizontal branch stars (temperature of about 8 keV).  The core temperature of supernova is of order
(30-70) MeV, which would allow for the production of $\phi$.  Once produced, $\phi$ will escape, carrying energy
with it.  This process should not be the dominant source of energy loss in supernova, as neutrinos with the expected
properties were detected from SN1987a~\cite{raffelt}.  The analysis of $\phi$ emission is similar to the case of Majorons for which the supernova bound on neutrino coupling is $\leq 4\times 10^{-7}$ \cite{farzan}. In our model, the scalar field coupling to electrons is $\sim \frac{m_e u}{v^2}\lambda_{H\phi}\simeq 10^{-18}$. It therefore easily satisfies this supernova bound.

\begin{figure}[h!]
\centering
\includegraphics[scale=0.5]{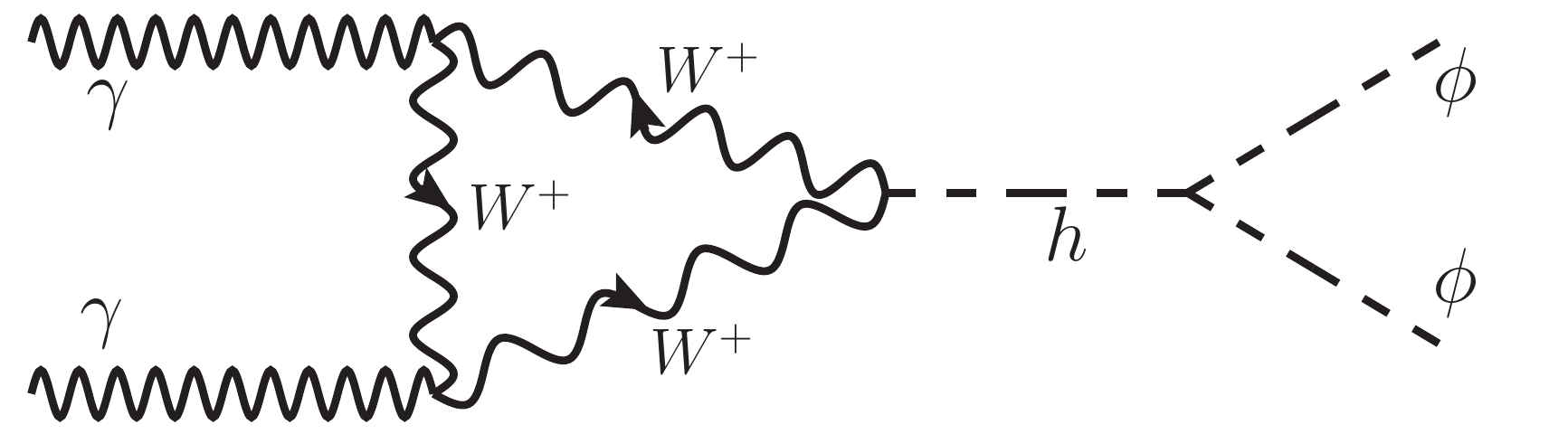}
\caption{Diagram responsible for $\phi$ pair creation inside supernovae.  There is another diagram where the $W$ boson inside
the loop is replaced by the top quark.}
\label{super}
\end{figure}

$\phi$ can also be pair produced inside supernova core via the process $\gamma \gamma \rightarrow \phi \phi$.  The
diagram responsible for this is shown in Fig. \ref{super}.  Here we make a rough estimate of the energy carried by $\phi$
and ascertain that this is indeed a negligible portion of the total energy of the supernova.  For a crude estimate,
we use the expression for the rate of energy loss $Q$,
\begin{equation}
Q = V_{\rm core} n_\gamma^2 \left\langle E \right \rangle \sigma(\gamma \gamma \rightarrow \phi \phi)~.
\end{equation}
Here $V_{\rm core} = 4 \pi R_{\rm core}^3/3$ is the core volume, and we take $R_{\rm core} = 10$ km. $n_\gamma$ is the
number density of photons, which we obtain from a thermal distribution.  The $\phi$ pair production cross section is
found to be
\begin{equation}
\sigma(\gamma \gamma \rightarrow \phi \phi) = \frac{\lambda_{H\Phi}^2v^2 E^2}{32 \pi M_h^4}\left(\frac{e^2g F}{16 \pi^2 m_W} \right)^2
\end{equation}
 where $F \simeq 6.54$ is the numerical value of the loop function associated with the decay of Higgs $h$ into two photons.
 We use $n_\gamma \simeq 0.2 T_\gamma^3$ and $E \simeq 3.15 T_\gamma$.  With these values, we find $Q \sim 3.6 \times 10^{54}
 \lambda_{H\Phi}^2$ erg/s, when $T_\gamma = 50$ MeV is used.  The supernova explosion lasted for about 10 seconds, so the total energy lost in $\phi$ would  be about ten times larger.  Demanding that the total energy lost is less than $10^{53}$ erg yields a very mild limit
 $|\lambda_{H\Phi}| < 5 \times 10^{-2}$. If  $T_\gamma = 70\, (100)$ MeV is used, this limit becomes 
 $|\lambda_{H\Phi}| < 1.1 \times 10^{-2}\, (2.3 \times
 10^{-3})$, which is easily satisfied in the model.  We note that all other energy loss processes in $\phi$ are highly suppressed.

\subsection{Explicit breaking of the reflection symmetry}

Here we analyze the consequences of explicitly breaking the reflection symmetry $\Phi \rightarrow -\Phi$ that was assumed in the preceding
discussions. If the model is embedded in a $U(1)$ gauge symmetry, as we do in the next section, the reflection symmetry is automatically present.  It is, however, interesting
to explore the modifications resulting from relaxing this symmetry. In this case, the following new terms in the Higgs potential are allowed,
in addition to the ones shown in Eq. (\ref{pot}):
\begin{equation}
V_{\rm new} = \mu_\Phi (\Phi-u)^3 + \mu_{H\Phi}\, (H^\dagger H - v^2) (\Phi -u)~.
\label{potnew}
\end{equation}
The $h-\phi$ mixing matrix will now become
\begin{eqnarray}
{\cal M}^2 = \left(\begin{matrix}\lambda_H v^2 & \sqrt{2} v\, (\lambda_{H\Phi}\, u + \mu_{H\Phi}) \\
\sqrt{2} v\, (\lambda_{H\Phi}\, u +\mu_{H\Phi}) & 2 \lambda_{\Phi} u^2 \end{matrix}\right)~.
\end{eqnarray}
The modified $H-\Phi$ mixing angle is
\begin{equation}
\theta_{H\Phi} \simeq \frac{\sqrt{2} v(\lambda_{H\Phi}\, u +\mu_{H\Phi})} {M_h^2}.
\end{equation}
The X-ray spectral anomaly then determines from $\phi \rightarrow \gamma \gamma$ decay rate
\begin{equation}
|\lambda_{H\Phi}\, u +\mu_{H\Phi}| = (3.6 \times 10^{-11} - 1.2 \times 10^{-11})~{\rm GeV}~.
\end{equation}

The requirement that $\phi$ never thermalizes sets a limit of $\mu_{H\Phi} < 8$ keV, obtained
from the process $hh \rightarrow \phi$ which has a rate $\Gamma \sim \mu_{H\Phi}^2/(8\pi M_h)$.
With this condition, the relic density of $\phi$ is unaltered from the calculation of Sec. \ref{relic}, and would
fix $\lambda_{H\Phi} \simeq 4.7 \times 10^{-9}$ as before.  The determination of the VEV $u$ of Eq. (\ref{uvalue})
will now be modified to
\begin{equation}
\pm u = (7.7 - 2.6)~ {\rm MeV} - \left(\frac{\mu_{H\Phi}}{4.7 \times 10^{-9}}\right)~.
\end{equation}
If $\mu_{H\Phi}$ is of order $0.1$ eV, $u$ will remain in the MeV range, and our discussions of earlier sections will
carry over.  This is what we shall assume in the remainder of this subsection, although larger values of $u$ and
$\mu_{H\Phi}$, as large as 100 GeV, are possible with a fine-tuning to get the $h-\phi$ mixing angle sufficiently small.

The self-interaction of dark matter is now modified, owing to the coupling $\mu_\Phi$ in Eq. (\ref{potnew}).  In
the nonrelativist limit, the cross section for $\phi \phi \rightarrow \phi \phi$ is given by
\begin{equation}
\sigma = \frac{9 \lambda_\Phi^2}{2\pi M_\phi^2} G(x),
\end{equation}
where
\begin{equation}
G(x) = 1+ 5 \sqrt{2} x + \frac{35}{2} x^2 + \frac{25}{\sqrt{2}} x^3 + \frac{25}{4} x^4
\end{equation}
with
\begin{equation}
x \equiv \frac{\mu_\Phi}{\sqrt{\lambda_\Phi} M_\phi}~.
\end{equation}
For positive values of $x$, $\sigma/M_\phi$ will exceed 1 barn/GeV, which is disfavored.  For
$-\sqrt{2} <  x < 0$ there is a reduction in this cross section relative to the case of exact reflection symmetry,
with $\sigma$ vanishing for
$x = -1.023$ and for $x=-0.391$.  Such a choice would correspond to $\mu_\Phi$ being of order eV.
These values of $x$ would be in full agreement with dark matter
self-interaction limits.

Finally, if the reflection symmetry is explicitly broken even by a small amount, stable domain walls will not
form \cite{goran}, and would relax the constraints resulting from there.

\section{Embedding in a simple hidden U(1) gauge model}

The model presented here has a natural embedding in a $U(1)$ gauge theory with a Higgs mechanism.
Such a model would be among the simplest gauge extensions of the Standard Model that could provide a rational for the small mass of the 7 keV scalar.  The gauge group of the SM is extended
to include an additional $U(1)_X$ which is spontaneously broken by a complex Higgs scalar field $\Phi$.  The new terms
in the Lagrangian are
\begin{equation}
{\cal L} = -\frac{1}{4} X_{\mu \nu}X^{\mu \nu} + \left|\partial_\mu \Phi -i g_X X_\mu \Phi \right|^2 - V(\Phi)~.
\end{equation}
The Higgs potential for the full SM plus $U(1)$ model now has the form
\begin{equation}
V = \frac{\lambda_H}{4} \left(H^\dagger H-v^2\right)^2 + \frac{\lambda_\Phi}{4} \left(|\Phi|^2-u^2\right)^2 + \frac{\lambda_{H\Phi}}{2}
\left(H^\dagger H-v^2\right)\left(|\Phi|^2-u^2\right)~.
\end{equation}

After spontaneous symmetry breaking with $\left\langle \Phi \right \rangle = u$, the phase of $\Phi$ is absorbed by the gauge
boson $X$ which becomes massive.  The left-over Higgs $\phi$ (real part of the complex field $\Phi$) will have exactly the same properties as the $\phi$ field
of the model that was
discussed in previous sections.  Owing to the existence of a new gauge boson $X$, there are some differences which we outline here.

It is easy to see that even after spontaneous symmetry breaking, there is an unbroken hidden parity symmetry in the model.
Under this symmetry only the $X$ gauge boson is odd while the $\phi$ field is even.  As a result, the $X$ gauge boson will be
stable and may contribute to the energy density of the Universe.  However, as we show below, this contribution is negligible.
A kinetic mixing term can break this symmetry and connect the ``dark $U(1)$" sector to the standard model particles.
  Actually, owing to this symmetry, if such a kinetic mixing term is not added,
it will never be induced in the theory. This is what we shall assume here.  If a kinetic mixing of the type $-(\zeta/2)\,B_{\mu \nu} X^{\mu \nu}$
is added to the Lagrangian, from the requirement that the hidden $X$ gauge boson never thermalizes down to temperatures of order MeV
sets a restriction $|\zeta| < 3 \times 10^{-10}$.

\begin{figure}[h!]
\centering
\includegraphics[scale=0.3]{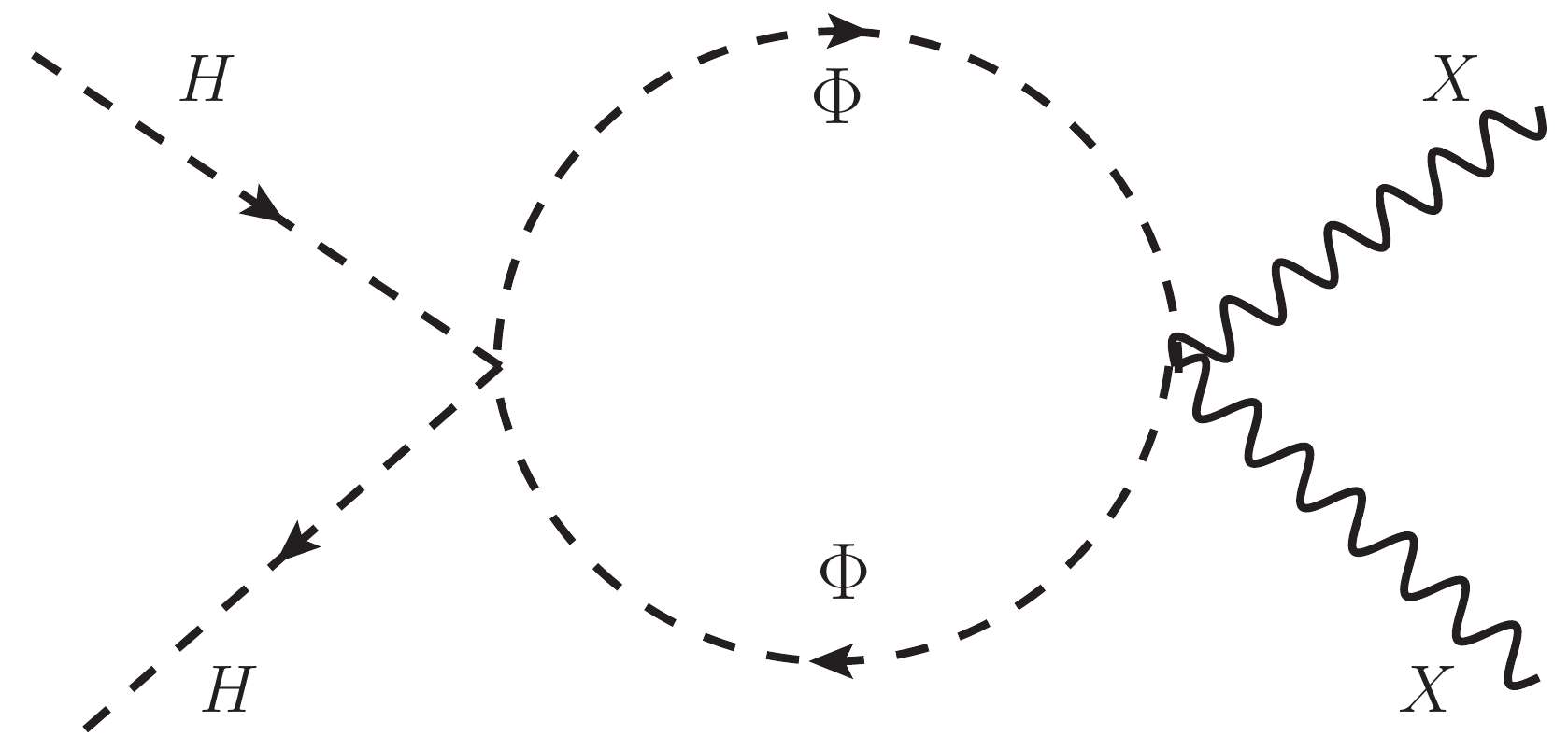} \vspace*{0.3in}
\includegraphics[scale=0.3]{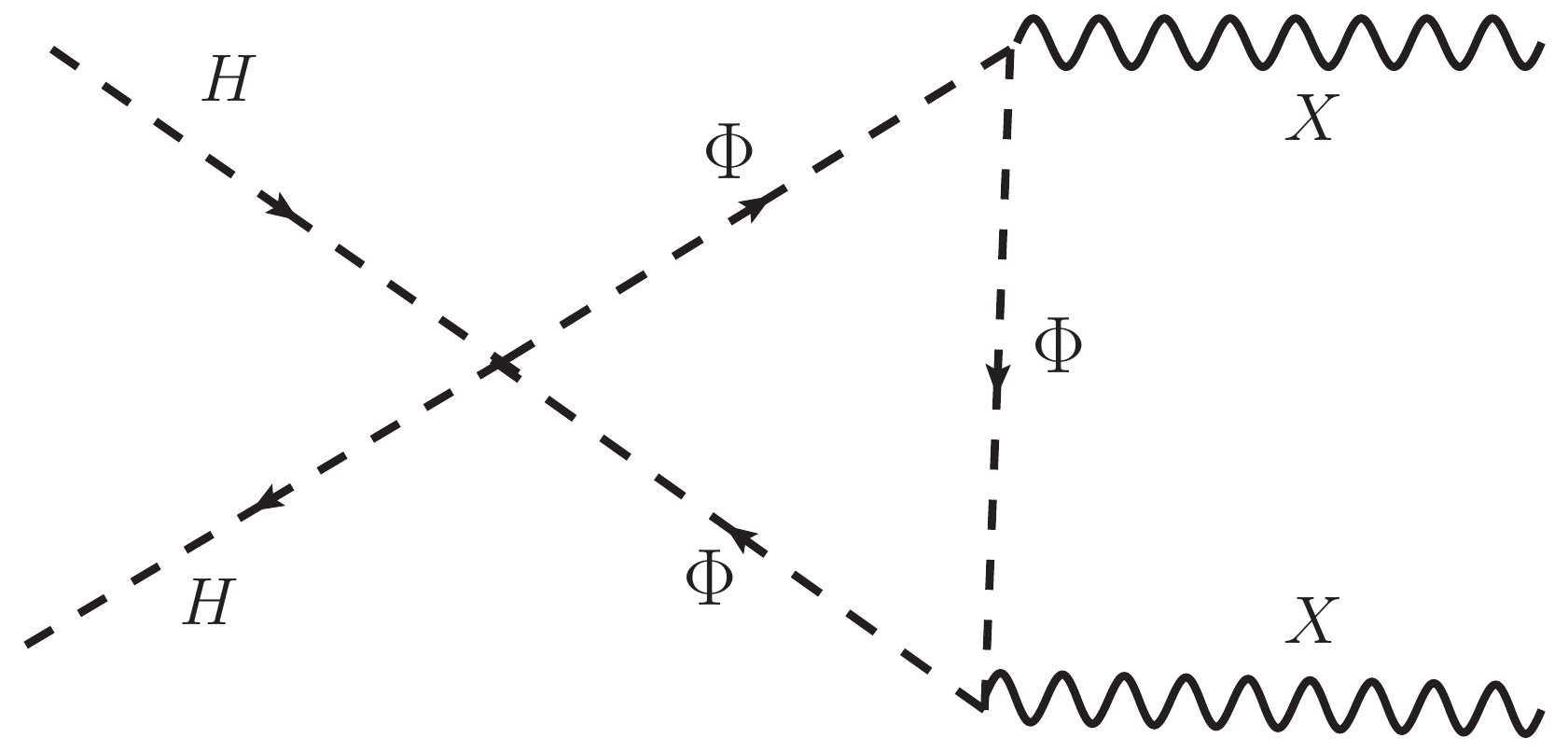}
\caption{Diagrams for production of $X$ gauge bosons in Higgs boson decays.}
\label{decay}
\end{figure}
As in the case of $\phi$ field, the dark $U(1)$ gauge boson never thermalizes with the plasma.  It has a mass given by $M_X = g_X u$
with $u$ listed as in Eq. (\ref{uvalue}).  $M_X$ should be larger than 7.1 keV, and because of the constraint from  Eq. (\ref{uvalue})
$M_X$ cannot exceed about 10 MeV.  $X$ can be produced in the early Universe in the decays and scattering of $h$ as shown in Fig. \ref{decay}.
Focussing on the decay, the width for $X$ production is given by
\begin{equation}
\Gamma(h \rightarrow XX )= \frac{\lambda_{H \Phi}^2 \alpha_X^2}{128\pi^3}\frac{v^2}{M_h}~.
\end{equation}
Comparing with the decay width for $h \rightarrow \phi \phi$, we find
\begin{equation}
\frac{\Gamma(h \rightarrow XX) }{\Gamma(h \rightarrow \phi \phi)} = \frac{\alpha_X^2}{4\pi^2}~.
\end{equation}
Consequently, the abundance of $X$ today is related to the abundance of $\phi$ as
\begin{equation}
\frac{\Omega_X}{\Omega_\phi} = \frac{\alpha_X^2}{4\pi^2} \frac{M_X}{M_\phi}~.
\end{equation}
If we demand this ratio to be $<1$, we obtain
\begin{equation}
\alpha_X < 0.5 \left(\frac{1~{\rm MeV}}{M_X} \right)^{1/2}
\end{equation}
which is always satisfied as long as the hidden $U(1)$ gauge coupling is in the perturbative range.
Actually, the hidden $U(1)$ gauge coupling $g_X$ should be naturally of order $0.1$, as it
would induce a quartic $\phi^4$ coupling with a coefficient $\lambda_\Phi \sim g_X^4/(16 \pi^2)$.
For $g_X = 0.1$, the induced value is $\lambda_\Phi \sim 10^{-6}$, which is consistent with the
dark matter phenomenology in the previous section.

We thus see that a natural embedding of the model of Sec. 2 into a local $U(1)$ gauge symmetry
preserves all the features that are desirable to explain the X-ray spectrum anomaly as well as
dark matter.

\section{Concluding remarks}

We conclude by making a few observations.

A light scalar particle can also lead to deviations from equivalence principle  at short distances~\cite{adel}. A 7 keV mass corresponds to range of force equal to $\sim 3\times 10^{-9}$ cm. At such short distances, there are no known bounds since it is not easy to probe such small distances by gravity experiments.

Photons can be produced via $\phi \phi \rightarrow \gamma \gamma$ scattering, in addition to the decay $\phi \rightarrow \phi \phi$.
However, the cross section times velocity for the scattering is found to be
$\sigma |v| \simeq 2 \times 10^{-53} \lambda_{H\Phi}^2$ cm$^2$, which is too small to be be observable in the X-ray spectrum.

%\end{itemize}

In summary, we have presented a very simple extension of the standard model by adding a real singlet scalar which can not only become a warm dark matter of the universe but also can explain the recently reported 3.55 keV X-ray lines in the sky as a result of dark matter decay.
The model has a natural embedding in a hidden $U(1)$ model with the Higgs mechanism.

\section*{Acknowledgement}
The work of KSB is supported in part by the US Department of Energy Grant No. de-sc0010108 and  RNM is supported
in part by the National Science Foundation Grant No. PHY-1315155.  We thank Saki Khan for numerical help.


\begin{thebibliography}{99}
\bibitem{Bulbul:2014sua}
  E.~Bulbul, M.~Markevitch, A.~Foster, R.~K.~Smith, M.~Loewenstein and S.~W.~Randall,
  %``Detection of An Unidentified Emission Line in the Stacked X-ray spectrum of Galaxy Clusters,''
  arXiv:1402.2301 [astro-ph.CO].
\bibitem{Boyarsky:2014jta}
  A.~Boyarsky, O.~Ruchayskiy, D.~Iakubovskyi and J.~Franse,
  %``An unidentified line in X-ray spectra of the Andromeda galaxy and Perseus galaxy cluster,''
  arXiv:1402.4119 [astro-ph.CO].
  \bibitem{scott}  S. Dodelson and L. M. Widrow, Phys. Rev. Lett. {\bf 72}, 17 (1994), hep-ph/9303287; X.-d. Shi and G. M. Fuller, Phys. Rev. Lett. {\bf 82}, 2832 (1999), astro-ph/9810076;  K. Abazajian, G. M. Fuller, and M. Patel, Phys. Rev. {\bf D64}, 023501 (2001), astro-ph/0101524; for a review, see A.~Kusenko,
  %``Sterile neutrinos: The Dark side of the light fermions,''
  Phys.\ Rept.\  {\bf 481}, 1 (2009)
  [arXiv:0906.2968 [hep-ph]].


\bibitem{Abazajian:2014gza}
  K.~N.~Abazajian, %``Resonantly-Produced 7 keV Sterile Neutrino Dark Matter Models and the Properties of Milky Way Satellites,''
  arXiv:1403.0954 [astro-ph.CO].

  \bibitem{axion}
   J.~Jaeckel, J.~Redondo and A.~Ringwald,
  %``A 3.55 keV hint for decaying axion-like particle dark matter,''
  arXiv:1402.7335 [hep-ph];
   M.~Cicoli, J.~P.~Conlon, M.~C.~D.~Marsh and M.~Rummel,
  %``A 3.55 keV Photon Line and its Morphology from a 3.55 keV ALP Line,''
  arXiv:1403.2370 [hep-ph];
  H.~M.~Lee, S.~C.~Park and W.~-I.~Park,
  %``Cluster X-ray line at $3.5\,{\rm keV}$ from axion-like dark matter,''
  arXiv:1403.0865 [astro-ph.CO]; K.~Nakayama, F.~Takahashi and T.~T.~Yanagida,
  %``Anomaly-free flavor models for Nambu-Goldstone bosons and the 3.5 keV X-ray line signal,''
  arXiv:1403.7390 [hep-ph].

  \bibitem{others}
  H.~Ishida, K.~S.~Jeong and F.~Takahashi,
  %``7 keV sterile neutrino dark matter from split flavor mechanism,''
  arXiv:1402.5837 [hep-ph];  T.~Higaki, K.~S.~Jeong and F.~Takahashi,
  %``The 7 keV axion dark matter and the X-ray line signal,''
  arXiv:1402.6965 [hep-ph]; D.~P.~Finkbeiner and N.~Weiner,
  %``An X-Ray Line from eXciting Dark Matter,''
  arXiv:1402.6671 [hep-ph];
  R.~Krall, M.~Reece and T.~Roxlo,
  %``Effective field theory and keV lines from dark matter,''
  arXiv:1403.1240 [hep-ph];
  C.~•m.~E.~Aisati, T.~Hambye and T.~Scarna,
  %``Can a millicharged dark matter particle emit an observable gamma-ray line?,''
  arXiv:1403.1280 [hep-ph];
  M.~Frandsen, F.~Sannino, I.~M.~Shoemaker and O.~Svendsen,
  %``X-ray Lines from Dark Matter: The Good, The Bad, and The Unlikely,''
  arXiv:1403.1570 [hep-ph];
    R.~Allahverdi, B.~Dutta and Y.~Gao,
  %``keV Photon Emission from Light Nonthermal Dark Matter,''
  arXiv:1403.5717 [hep-ph];
  C.~Kolda and J.~Unwin,
  %``X-ray lines from R-parity violating decays of keV sparticles,''
  arXiv:1403.5580 [hep-ph];
  S.~P.~Liew,
  %``Axino dark matter in light of an anomalous X-ray line,''
  arXiv:1403.6621 [hep-ph];
  N.~-E.~Bomark and L.~Roszkowski,
  %``The 3.5 keV X-ray line from decaying gravitino dark matter,''
  arXiv:1403.6503 [hep-ph]; K.~-Y.~Choi and O.~Seto,
  %``X-ray line signal from decaying axino warm dark matter,''
  arXiv:1403.1782 [hep-ph]; S.~Baek and H.~Okada,
  %``7 keV Dark Matter as X-ray Line Signal in Radiative Neutrino Model,''
  arXiv:1403.1710 [hep-ph]; P.~Ko, Z.~kang, T.~Li and Y.~Liu,
  %``Natural $X$-ray Lines from the Low Scale Supersymmetry Breaking,''
  arXiv:1403.7742 [hep-ph];
 S.~V.~Demidov and D.~S.~Gorbunov,
  %``SUSY in the sky or keV signature of sub-GeV gravitino dark matter,''
  arXiv:1404.1339 [hep-ph];
  F.~S.~Queiroz and K.~Shinha,
  %``The Poker Face of the Majoron Dark Matter Model: LUX to keV Line,''
  arXiv:1404.1400 [hep-ph]; %E.~Dudas, L.~Heurtier and Y.~Mambrini,
  %``Generating X-ray lines from annihilating dark matter,''
  %arXiv:1404.1927 [hep-ph];
  K. Kong, J.-C. Park and S.C. Park,
%``X-ray line signal from 7 keV axino dark matter decay,''
[arXiv:1403.1536 [hep-ph]]; A.~G.~Dias, A.~C.~B.~Machado, C.~C.~Nishi, A.~Ringwald and P.~Vaudrevange,
  %``The Quest for an Intermediate-Scale Accidental Axion and Further ALPs,''
  arXiv:1403.5760 [hep-ph].

  \bibitem{seckel}
  I.~Z.~Rothstein, K.~S.~Babu and D.~Seckel,
  %``Planck scale symmetry breaking and majoron physics,''
  Nucl.\ Phys.\ B {\bf 403}, 725 (1993)  [hep-ph/9301213].


  \bibitem{real}  V.~Silveira and A.~Zee,
  %``Scalar Phantoms,''
  Phys.\ Lett.\ B {\bf 161}, 136 (1985); J.~McDonald,
  %``Gauge singlet scalars as cold dark matter,''
  Phys.\ Rev.\ D {\bf 50}, 3637 (1994).

  \bibitem{real1} For phenomenological studies of ``SM+real singlet" models, see  M. C. Bento, O. Bertolami, R. Rosenfeld, and L. Teodoro, Phys. Rev. D62, 041302 (2000),
astro-ph/0003350; C. P. Burgess, M. Pospelov, and T. ter Veldhuis, Nucl. Phys. B619, 709 (2001), hep-
ph/0011335; H. Davoudiasl, R. Kitano, T. Li, and H. Murayama, Phys. Lett. B609, 117 (2005), hep-
ph/0405097; V.~Barger, P.~Langacker, M.~McCaskey, M.~J.~Ramsey-Musolf and G.~Shaughnessy,
  %``LHC Phenomenology of an Extended Standard Model with a Real Scalar Singlet,''
  Phys.\ Rev.\ D {\bf 77}, 035005 (2008);  X.~-G.~He, S.~-Y.~Ho, J.~Tandean and H.~-C.~Tsai,
  %``Scalar Dark Matter and Standard Model with Four Generations,''
  Phys.\ Rev.\ D {\bf 82}, 035016 (2010)
  [arXiv:1004.3464 [hep-ph]].

  \bibitem{gondolo}
   P.~Gondolo and G.~Gelmini,
  %``Cosmic abundances of stable particles: Improved analysis,''
  Nucl.\ Phys.\ B {\bf 360}, 145 (1991).

  \bibitem{dolan}
   L.~Dolan and R.~Jackiw,
  %``Symmetry Behavior at Finite Temperature,''
  Phys.\ Rev.\ D {\bf 9}, 3320 (1974).  %%CITATION = PHRVA,D9,3320;%%

\bibitem{okun} Y.~B.~Zeldovich, I.~Y.~Kobzarev and L.~B.~Okun,
  %``Cosmological Consequences of the Spontaneous Breakdown of Discrete Symmetry,''
  Zh.\ Eksp.\ Teor.\ Fiz.\  {\bf 67}, 3 (1974)
  [Sov.\ Phys.\ JETP {\bf 40}, 1 (1974)];  for reviews and references, see A.~Vilenkin,
  %``Cosmic Strings and Domain Walls,''
  Phys.\ Rept.\  {\bf 121}, 263 (1985); T.~Vachaspati,
  ``{\it Kinks and domain walls: An introduction to classical and quantum solitons,}''
  Cambridge, UK: Univ. Pr. (2006) 176 p.

  \bibitem{self} M.~Kaplinghat, R.~E.~Keeley, T.~Linden and H.~-B.~Yu,
  %``Tying Dark Matter to Baryons with Self-interactions,''
  arXiv:1311.6524 [astro-ph.CO]; M.~Kaplinghat, S.~Tulin and H.~-B.~Yu,
  %``Self-interacting Dark Matter Benchmarks,''
  arXiv:1308.0618 [hep-ph].


 \bibitem{raffelt} For studies of supernova energy loss constraints on light particles, see
  R.~Barbieri and R.~N.~Mohapatra,
  %``Limits on Right-handed Interactions From {SN1987A} Observations,''
  Phys.\ Rev.\ D {\bf 39}, 1229 (1989); G. Raffelt, ``{\it Stars as Laboratories for Fundamental Physics}", Chicago University Press (1999).


\bibitem{farzan} For a recent discussion and earlier references, see Y.~Farzan,
  %``Bounds on the coupling of the Majoron to light neutrinos from supernova cooling,''
  Phys.\ Rev.\ D {\bf 67}, 073015 (2003)
  [hep-ph/0211375].

  \bibitem{goran}

   B.~Rai and G.~Senjanovic,
  %``Gravity and domain wall problem,''
  Phys.\ Rev.\ D {\bf 49}, 2729 (1994)  [hep-ph/9301240].

   \bibitem{adel} For a summary of these bounds, see C.~D.~Hoyle, U.~Schmidt, B.~R.~Heckel, E.~G.~Adelberger, J.~H.~Gundlach, D.~J.~Kapner and H.~E.~Swanson,
  %``Submillimeter tests of the gravitational inverse square law: a search for 'large' extra dimensions,''
  Phys.\ Rev.\ Lett.\  {\bf 86}, 1418 (2001).








\end{thebibliography}
\end{document}